\documentclass[prb,twocolumn]{revtex4b4}
\usepackage{graphicx}
\begin{document}
\title{Global Optimization and the Energy Landscapes of Dzugutov Clusters}
\author{Jonathan P.~K.~Doye, David J.~Wales}
\affiliation{University Chemical Laboratory, Lensfield Road, Cambridge CB2 1EW, United Kingdom}
\author{Sergei I.~Simdyankin}
\affiliation{Department of Numerical Analysis, Royal Institute of Technology, 
SE-100 44 Stockholm, Sweden}
\date{\today}
\begin{abstract}
The global minima of clusters bound by a Dzugutov potential form non-compact
polytetrahedral clusters mainly composed of interpenetrating and 
face-sharing 13-atom icosahedra.
As the size increases, these icosahedral units first form linear arrays, 
then two-dimensional rings, then three-dimensional networks.
Characterization of the energy landscapes of these clusters shows that 
they are particularly rough and generally exhibit a multiple-funnel topography. 
These results provide new insights into the structure and dynamics of
bulk supercooled Dzugutov liquids and the form of the bulk phase diagram.
\end{abstract}
\maketitle

\section{Introduction}

One reason for the continued growth in cluster science is the link that these finite systems provide 
between atoms and molecules and bulk systems. 
Indeed, much cluster research has focussed on the size 
evolution of cluster properties in order to understand how bulk behaviour, such as phase transitions and 
crystalline structure, emerge with increasing size. Furthermore, from this research new insights
into bulk behaviour can often be obtained. 

For example, the structure of clusters
has often been invoked when considering the local structure of bulk liquids and glasses.
Frank may have been the first to make this connection.\cite{Frank52} 
He noted that the 13-atom icosahedron has a lower energy than a close-packed cluster for the 
Lennard-Jones (LJ) potential. From this result he inferred that the most favoured structure of
liquids would involve icosahedral local ordering, thus making homogeneous nucleation of a crystal 
from a liquid difficult, because crystal formation would require a change in the local order.\cite{Frankbug}
Since then convincing evidence that simple liquids have polytetrahedral\cite{ptet} order 
has been obtained,\cite{NelsonS} and so the use of clusters as models of 
local order in liquids has become common.\cite{Hoare76}
Furthermore, clusters with five-fold symmetry are often used 
as building blocks for quasicrystalline structures\cite{Elser85,Audier86,Stephens86,Borodin} 
and metallic crystals.\cite{Shoemaker}

Much of the computational work on supercooled liquids and glasses has focussed 
on the behaviour exhibited by systems interacting according to simple model potentials. 
One of the most commonly studied systems employs a binary LJ potential,\cite{Kob95}
where the binary nature of the system hinders the crystallization that readily occurs for
a single component LJ liquid. However, recent evidence suggests that the low
energy configurations for this system involve some demixing and crystalline ordering.\cite{Middleton00}
Furthermore, although this system is a good generic glass former, 
it does not capture all of the features associated with metallic glasses, 
for which particularly strong local polytetrahedral and icosahedral ordering occurs.

An alternative approach is to use a single component system with a potential that is
designed to prevent crystallization. This line of enquiry was pioneered by
Dzugutov, who introduced a potential with a local maximum at approximately $\sqrt 2$ times the 
equilibrium pair distance in order to disfavour close-packed structures.\cite{Dzugutov91,Dzugutov92} 
This maximum also bears a resemblance to the first of the Friedel oscillations 
that can occur for metal potentials. Systems interacting according to this Dzugutov potential
have been found to be good glass-formers, and to exhibit a first sharp diffraction
peak and a split second peak in the structure factor,\cite{Sadigh99} which are common features
of metallic glasses. One other interesting feature of this potential is that 
under certain conditions it can produce a dodecagonal quasicrystal.\cite{Dzugutov93}

Our aim here is to further understand the structural implications of the Dzugutov potential
by identifying the characteristic structural motifs associated with the global minima of Dzugutov clusters.
Furthermore, we examine the consequences of the potential for the topography of the
energy landscape [or potential energy surface (PES)] of these clusters. 
As well as the insights into bulk behaviour that these results provide,
they will also be potentially helpful in understanding the structures of clusters that favour polytetrahedral structure. 
Recent results on small cobalt clusters indicate that they are polytetrahedral,\cite{Dassenoy00}
although the detailed structure is unknown. Furthermore, one would expect clusters of alloys that
adopt a Frank-Kasper structure\cite{Shoemaker,FrankK58,FrankK59} in bulk to show polytetrahedral order in clusters.

\section{Methods}

\subsection{Potentials}

The Dzugutov pair potential is:\cite{Dzugutov91,Dzugutov92}
\begin{eqnarray}
V(r)&=&A(r^{-m}-B) \exp\left( {c\over r-a}\right) \Theta(a-r) + \nonumber \\
    & &B \exp\left({d\over r-b}\right) \Theta(b-r), 
\end{eqnarray} 
where $\Theta(x)$, the Heaviside step function, is 0 for $x<0$ and 1 otherwise, 
and the parameters in the potential have the values
\begin{eqnarray}
A=5.82& \quad &a=1.87 \nonumber \\
B=1.28&  &b=1.94 \nonumber \\
c=1.10&  &d=0.27 \nonumber \\
m=16.&&
\end{eqnarray} 
The total potential energy of a cluster of Dzugutov atoms is then $E=\sum_{i<j} V(r_{ij})$, where
$r_{ij}$ is the distance between atoms $i$ and $j$.
The pair potential has a minimum at $r_{\rm eq}$=1.130  of depth $\epsilon$=$0.581$ and 
a maximum at $r_{\rm max}$=1.628=1.44$\,r_{\rm eq}$ of height 0.460 ($0.791\epsilon)$.

\begin{figure}
\includegraphics[width=8.2cm]{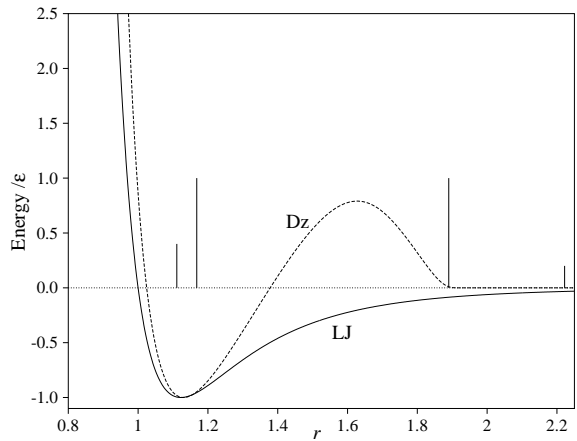}
\caption{\label{fig:potential} 
A comparison of the Dzugutov (Dz) and Lennard-Jones (LJ) potentials.
The unit of energy is the pair well depth. 
The vertical lines show the positions of the pair distances 
in the 13-atom icosahedron and their heights are proportional to the numbers of 
pairs with that distance.
}
\end{figure}

In Figure \ref{fig:potential} we compare the Dzugutov and the LJ potentials. 
The curvature at the bottom of the Dzugutov well in the above units is virtually 
the same as that of the LJ potential in reduced units. 
However, when the unit of energy for both potentials is the pair well depth, as in Figure 
\ref{fig:potential}, the curvature at
the minimum of the Dzugutov potential is somewhat larger,
i.e.\ the Dzugutov potential well is narrower. 

This effect can be quantified.
In units of pair well depth and equilibrium pair separation, the curvature at the minimum 
of the Morse potential 
\begin{equation}
V(r)/\epsilon=\exp\left[2\rho\left(1-{r\over r_{\rm eq}}\right)\right]-
             2\exp\left[\rho\left(1-{r\over r_{\rm eq}}\right)\right],
\end{equation}
is 2$\rho^2$ and of the generalized LJ potential 
\begin{equation}
V(r)=4\epsilon \left[\left({r\over\sigma}\right)^{2n}-\left({r\over\sigma}\right)^n\right] 
\end{equation}
is 2$n^2$.
These results allow one to calculate an effective $n$ (or $\rho$) value. 
For the Dzugutov potential
$n_{\rm eff}^{\rm Dz}$=$\rho_{\rm eff}^{\rm Dz}$=7.52, as compared to $n^{\rm LJ}$=6.
This comparison will be important to our interpretation of the structures of these clusters
since, as illustrated for Morse clusters,\cite{Doye95c,Doye97d} the well width
is a key factor in determining a cluster's structure.

For a pair potential we can partition the energy into three terms:\cite{Doye95c}
\begin{equation}
\label{eq:Esplit}
E=-n_{\rm nn}\epsilon+E_{\rm strain}+E_{\rm nnn},
\end{equation}
where 
$\epsilon$ is the pair well depth, 
$n_{\rm nn}$ is the number of nearest neighbours (two atoms
are defined as nearest neighbours if $r_{ij}<r_0$),
\begin{eqnarray}
E_{\rm strain}&=&\sum_{i<j,r_{ij}<r_0} (V(r_{ij})+\epsilon) \quad \hbox{  and  } \\
E_{\rm nnn}&=&\sum_{i<j,r_{ij}\ge r_0} V(r_{ij}).
\end{eqnarray}
We choose $r_0$=1.378 where $V(r_0)$=0. 
The first term in Equation (\ref{eq:Esplit}) is the ideal pair energy if all
$n_{\rm nn}$ nearest-neighbour pairs lie exactly at the equilibrium pair distance.
The strain energy, $E_{\rm strain}$, is the energetic penalty for the deviation of 
nearest-neighbour distances from the equilibrium pair distance, 
and $E_{\rm nnn}$ is the contribution to the energy from
non-nearest neighbours. For the Dzugutov potential $E_{\rm nnn}$ corresponds to the
energetic penalty for distances lying near to $\sqrt 2\, r_{\rm eq}$ (more precisely 
in the range $r_0<r<b$). 
The global minimum of a cluster represents the best balance between maximizing 
$n_{\rm nn}$ whilst minimizing both the strain energy and $E_{\rm nnn}$.

The big difference between the Dzugutov potential and most other pair potentials is 
the local maximum at $r_{\rm max}$ and the consequent positive value of the term $E_{\rm nnn}$.
This term disfavours distances near to $\sqrt 2\, r_{\rm eq}$ that are associated 
with the octahedral interstices in close-packed structures. For clusters this term
also disfavours Marks decahedra\cite{Marks84} and Mackay\cite{Mackay} icosahedra with more than one shell,
because the strained face-centred cubic (fcc) tetrahedra that make up these structures contain octahedral 
interstices. However, as Figure \ref{fig:potential} shows, the distances in the 13-atom
icosahedron are able to avoid this maximum. 
Therefore, $E_{\rm nnn}$ favours polytetrahedral clusters. 
For larger polytetrahedral clusters the value of $E_{\rm nnn}$ is somewhat correlated with $E_{\rm strain}$. 
Strain results from a dispersion of nearest-neighbour distances, and so is also likely to lead to
a greater dispersion in the second nearest-neighbour distances 
and potentially more distances that stray towards the maximum at $r_{\rm max}$. 

We have previously analysed the roles of $n_{\rm nn}$ and $E_{\rm strain}$ for 
other pair potentials, such as the Morse potential.\cite{Doye95c,Doye97d}
A large value of $n_{\rm nn}$ is achieved by having a compact spherical shape with
a large proportion of surfaces with high coordination numbers, e.g.\ fcc \{111\} faces.
$E_{\rm strain}$  disfavours strained structures and the magnitude of this term 
increases as the potential well becomes narrower. The strain for close-packed 
structures can be zero, but increases along the series: decahedral, icosahedral and polytetrahedral.
Thus, the strain associated with compact polytetrahedral structures can be only accommodated 
at low $n_{\rm eff}$ values.
Therefore, given the value of $n_{\rm eff}^{Dz}$, compact polytetrahedral structures
seem unlikely and it is not self-evident what types of structure are best able to 
satisfy these competing demands.

\subsection{Global Optimization}

To obtain putative global minima for the Dzugutov clusters we used the basin-hopping\cite{WalesD97} 
(or Monte Carlo plus minimization\cite{Li87a}) algorithm, 
which has proved very useful in locating cluster global minima.\cite{WalesS99} 
In this algorithm a constant temperature Metropolis Monte Carlo simulation is
performed with the acceptance criterion based not upon the energy of the new configuration,
but upon the energy of the minimum obtained by a local minimization from
that configuration. Details of the basic approach employed in the present work may be
found elsewhere.\cite{WalesD97}

Dzugutov clusters are particularly challenging cases for global
optimization, and so for each cluster size five runs of 100\,000 steps each starting
from a random configuration were performed.
Short runs of 10\,000 steps were also performed for each cluster starting from the lowest-energy
minima found for the size above and below with an atom removed and added, respectively.
In the latter runs the seed geometries were fixed for the first 2\,000 steps.
These short runs were particularly important at larger sizes,
and were repeated until there were no further improvements for any size.
In each run a constant temperature of $0.4$ in reduced units was employed, with a fixed
maximum step size for Cartesian displacements of $0.36$. Random angular moves were performed
for the most weakly bound atom if its pair energy exceeded $3/5$ of the lowest pair energy 
for the structure in question. 
(The pair energy of atom $\alpha$ is defined as $\sum_{i\not=\alpha} V(r_{i\alpha})$.) 

Problems arise in the energy minimizations if atoms evaporate from the cluster:
there are no attractive forces on an atom lacking any neighbours within a distance of $r_{\rm max}$.
As in previous work we constrained the atoms in a sphere to limit the maximum interatomic distances.
However, since some global minima of the Dzugutov potential exhibit highly non-spherical
geometries it was necessary to enlarge the container to twice the radius used in previous work.\cite{WalesD97}
Atoms with pair energies of zero were therefore shifted towards the origin through a distance
of one unit on every call of the potential subroutine.

\subsection{Searching the potential energy surface}

To map the PES topography of some of the Dzugutov clusters we used the same methods as those
we have previously applied to LJ\cite{Doye99f,Doye99c,Doye00c} and Morse\cite{Miller99a} clusters.
We thereby obtain large samples of connected minima and transition states that provide
good representations of the low-energy regions of the PES.
The approach involves repeated applications of eigenvector-following\cite{Cerjan} to
find new transition states and the minima they connect, as described in detail elsewhere.\cite{WalesDMMW00}

\begin{table*}
\caption{\label{table:gmin} Energies and point groups of putative Dz$_N$ global minima.}
\begin{ruledtabular}
\begin{tabular}{ccccccccccc}
$N$  & PG & Energy &\ & $N$ & PG & Energy &\ & $N$ & PG & Energy \\
\hline
 3 & $D_{3h}$ &  $-1.744308$ & & 36 & $C_{2v}$ &  $-73.059795$ & &  69 & $C_1$    & $-147.488509$ \\
 4 & $T_d$    &  $-3.488617$ & & 37 & $D_{5d}$ &  $-75.473295$ & &  70 & $C_s$    & $-149.924691$ \\
 5 & $D_{3h}$ &  $-5.194997$ & & 38 & $D_{6h}$ &  $-78.212691$ & &  71 & $C_1$    & $-151.775786$ \\
 6 & $C_{2v}$ &  $-6.895466$ & & 39 & $C_s$    &  $-80.028043$ & &  72 & $C_1$    & $-154.077804$ \\
 7 & $D_{5h}$ &  $-9.110478$ & & 40 & $C_1$    &  $-82.007765$ & &  73 & $C_1$    & $-156.300056$ \\
 8 & $C_s$    & $-10.800965$ & & 41 & $C_s$    &  $-84.193592$ & &  74 & $C_1$    & $-158.621949$ \\
 9 & $C_{2v}$ & $-13.004336$ & & 42 & $C_{2v}$ &  $-86.888348$ & &  75 & $C_s$    & $-161.127938$ \\
10 & $C_{3v}$ & $-15.182299$ & & 43 & $C_1$    &  $-88.707201$ & &  76 & $D_{3d}$ & $-163.427645$ \\
11 & $C_{2v}$ & $-17.354191$ & & 44 & $C_s$    &  $-91.021780$ & &  77 & $C_s$    & $-165.950395$ \\
12 & $C_{5v}$ & $-19.973962$ & & 45 & $C_2$    &  $-92.914141$ & &  78 & $C_1$    & $-168.012336$ \\
13 & $I_h$    & $-23.206834$ & & 46 & $C_{2v}$ &  $-95.600895$ & &  79 & $C_s$    & $-170.278007$ \\
14 & $C_{3v}$ & $-24.800150$ & & 47 & $C_s$    &  $-97.424527$ & &  80 & $C_s$    & $-172.684322$ \\
15 & $C_{2v}$ & $-26.892331$ & & 48 & $C_1$    &  $-99.877799$ & &  81 & $C_1$    & $-174.703554$ \\
16 & $C_s$    & $-28.935002$ & & 49 & $C_1$    & $-101.992888$ & &  82 & $C_1$    & $-177.379501$ \\
17 & $C_s$    & $-31.071373$ & & 50 & $C_1$    & $-104.366190$ & &  83 & $C_1$    & $-179.785948$ \\
18 & $C_s$    & $-33.388902$ & & 51 & $C_1$    & $-106.628459$ & &  84 & $C_1$    & $-182.318738$ \\
19 & $D_{5h}$ & $-36.387297$ & & 52 & $C_2$    & $-109.182997$ & &  85 & $C_2$    & $-185.035277$ \\
20 & $C_{2v}$ & $-38.185980$ & & 53 & $C_1$    & $-111.160304$ & &  86 & $D_3$    & $-188.011095$ \\
21 & $C_1$    & $-40.164779$ & & 54 & $C_1$    & $-113.571440$ & &  87 & $C_1$    & $-190.060568$ \\
22 & $C_s$    & $-42.232479$ & & 55 & $C_s$    & $-115.796712$ & &  88 & $C_2$    & $-192.537449$ \\
23 & $D_{3h}$ & $-45.097045$ & & 56 & $C_s$    & $-118.113023$ & &  89 & $C_1$    & $-194.585540$ \\
24 & $C_s$    & $-46.967533$ & & 57 & $T_d$    & $-120.619381$ & &  90 & $C_2$    & $-197.065663$ \\
25 & $C_{2v}$ & $-49.351951$ & & 58 & $C_{2h}$ & $-122.721838$ & &  91 & $C_1$    & $-199.110106$ \\
26 & $C_s$    & $-51.162031$ & & 59 & $C_s$    & $-125.168480$ & &  92 & $D_3$    & $-201.595867$ \\
27 & $C_{2v}$ & $-53.321668$ & & 60 & $C_s$    & $-127.515076$ & &  93 & $C_1$    & $-203.620819$ \\
28 & $C_{2v}$ & $-55.482764$ & & 61 & $C_1$    & $-129.438766$ & &  94 & $C_1$    & $-205.650655$ \\
29 & $C_s$    & $-58.091939$ & & 62 & $C_s$    & $-131.779964$ & &  95 & $C_1$    & $-207.981315$ \\
30 & $C_1$    & $-59.907899$ & & 63 & $C_1$    & $-133.838716$ & &  96 & $C_1$    & $-210.175003$ \\
31 & $D_{5h}$ & $-62.387189$ & & 64 & $C_1$    & $-136.236101$ & &  97 & $C_1$    & $-212.509621$ \\
32 & $C_s$    & $-64.064478$ & & 65 & $C_s$    & $-138.685418$ & &  98 & $C_1$    & $-214.553532$ \\
33 & $D_{3d}$ & $-66.893638$ & & 66 & $C_s$    & $-141.219450$ & &  99 & $C_1$    & $-216.884240$ \\
34 & $C_1$    & $-68.696811$ & & 67 & $D_{2d}$ & $-143.476004$ & & 100 & $C_1$    & $-219.523265$ \\
35 & $C_2$    & $-71.134071$ & & 68 & $C_1$    & $-145.565041$ & & \\
\end{tabular}
\end{ruledtabular}
\end{table*}

\section{Results}

\subsection{Global Minima}

The lowest-energy minima that we found are given in Table \ref{table:gmin} up to $N$=100;
these results will also be made available in the Cambridge Cluster Database.\cite{Web}
For most clusters up to $N\sim 60$ these structures were obtained from basin-hopping runs
that were started from a random configuration. 
However, above this size we often found
that the lowest-energy minimum was only located in the runs seeded with a configuration
derived from the best structure at a smaller or a larger size. Therefore, it would not be surprising
if lower-energy minima exist for some of the larger clusters. 
However, any such minima would not be expected to change the observed structural motifs, which are 
the main focus of the present section.

\begin{figure}
\includegraphics[width=8.2cm]{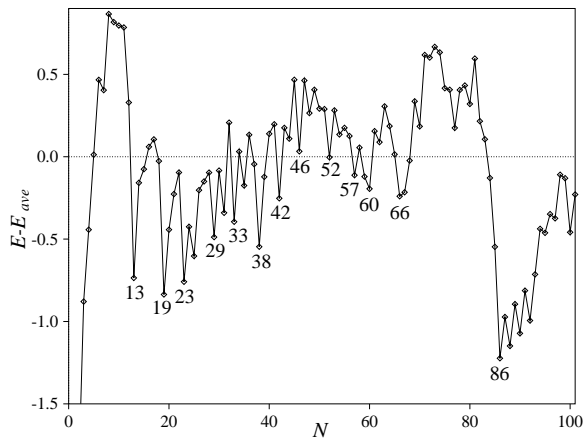}
\caption{\label{fig:EvN} 
The energies of the putative Dz$_N$ global minima measured with respect
to $E_{\rm ave}$, where $E_{\rm ave}=-2.605 N+ 2.568 N^{2/3}-4.859 N^{1/3}+8.614$. 
}
\end{figure}

\begin{figure}
\includegraphics[width=8.2cm]{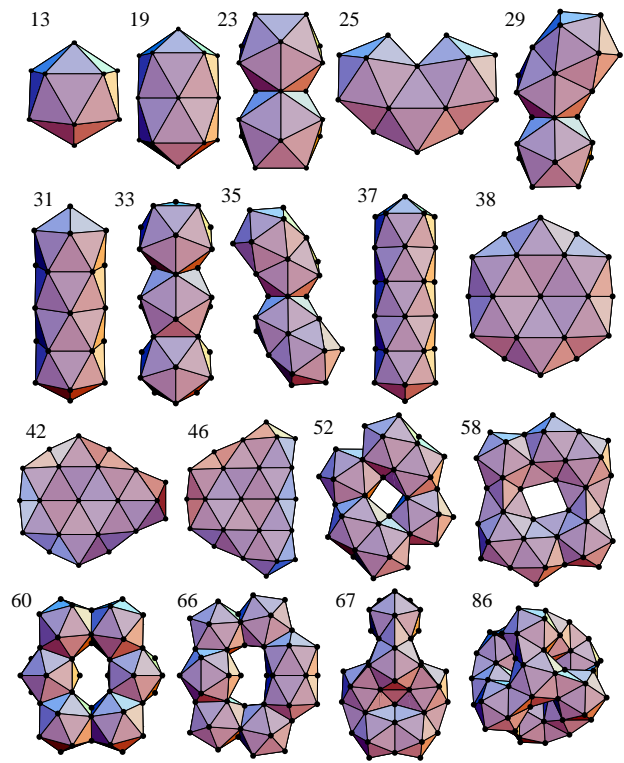}
\caption{\label{fig:structures} Some representative Dz$_N$ global minima.
The labels indicate the number of atoms in each cluster.
}
\end{figure}

The energies of the putative global minima are plotted in Figure \ref{fig:EvN} relative to the smooth function, 
$E_{\rm ave}$, which represents the average energy of the global minima, so that particularly 
stable sizes are readily apparent.
A representative set of global minima is depicted in Figure \ref{fig:structures}; 
most of these structures correspond either to a minimum in Figure \ref{fig:EvN} or a high symmetry configuration.

As noted earlier, one consequence of the maximum in the potential is that
polytetrahedral structures are likely to be favoured, 
and for small sizes the clusters adopt the expected structures.
The 13-atom global minimum is the icosahedron (Figure \ref{fig:structures}) 
and the global minima at smaller sizes lie on a polytetrahedral growth 
sequence to this structure.
This sequence is similar to the global minima for the LJ potential,\cite{HoareP71}
except that the Dz$_6$ global minimum is the bicapped tetrahedron, rather than
the octahedron.

There are two possible overlayers for growth on the 13-atom icosahedron that are frequently observed.
The first involves adding atoms to the faces and vertices of the icosahedron and
is termed the anti-Mackay overlayer. This overlayer continues the polytetrahedral
packing of the icosahedron and leads to a series of interpenetrating icosahedra.
The overlayer is completed at a 45-atom rhombic tricontahedron, which is an
icosahedron of interpenetrating icosahedra. The second possible overlayer involves adding atoms
to the edges and vertices of the icosahedron, and is termed the Mackay overlayer
because it leads to the 55-atom Mackay icosahedron. This overlayer introduces
octahedral interstices and so is disfavoured by the Dzugutov potential. No Dz$_N$ global
minima have this structure.

The global minima above $N$=13 initially have an anti-Mackay structure, for example,
the double icosahedron at $N$=19 (Figure \ref{fig:structures}).
This trend continues up to $N$=22. For $N$=23 the usual anti-Mackay structure is a 
triangular arrangement of three interpenetrating icosahedra. However the Dz$_{23}$ global minimum
consists of two face-sharing icosahedra (Figure \ref{fig:structures}), because the
anti-Mackay structure has a larger strain energy (4.810 as opposed to 3.138 for the global
minimum) and a slightly higher $E_{\rm nnn}$, making it 0.156 higher in energy, even though
it has three more nearest-neighbour contacts.

This trend is continued for larger sizes. 
The strain associated with compact polytetrahedral structures 
cannot be accommodated without too great an energetic penalty because of the 
value of $n_{\rm eff}^{\rm Dz}$.
Instead, polytetrahedral structures with extended configurations are favoured even though
they have fewer nearest neighbours.
These structures involve a combination of face-sharing and interpenetrating icosahedra 
(Figure \ref{fig:structures}). 
At $N$=25 a bent arrangement of three interpenetrating icosahedra is the global minimum
(it is just lower in energy than the linear triple icosahedron). 
At $N$=31 and 37 the global minima are linear quadruple and quintuple icosahedra.
At $N$=33 the global minimum is a linear arrangement of three face-sharing icosahedra.
At $N$=29 the global minimum involves an icosahedron sharing a face with
a double icosahedron and at $N$=35 it is two face-sharing double icosahedra. 

Above $N$=37 the global minima become two-dimensional in character. 
The 38-atom global minimum is a ring of six interpenetrating icosahedra. 
The nearest-neighbour contact along the six-fold axis of this structure forms 
the common edge for six tetrahedra, whereas all other internal nearest-neighbour contacts
are surrounded by five tetrahedra. 
This edge is said to have a disclination running along it,\cite{NelsonS} 
and it is not possible to form a bulk compact polytetrahedral structure 
without introducing such disclination lines.
Frank-Kasper phases\cite{FrankK58,FrankK59} and quasicrystals can be characterized by 
their networks of disclination lines.\cite{NelsonS}
Most of the clusters immediately above $N$=38 have structures based upon the Dz$_{38}$ global minimum, 
such as the 42- and 46-atom structures shown in Figure \ref{fig:structures}.

At larger sizes the global minima are rings of face-sharing icosahedra and double icosahedra 
that have a hole in the middle of the structure, e.g.\ $N$=52, 58, 60 and 66 in Figure \ref{fig:structures}.
Sometimes, as for $N=58$, not all the icosahedral units are face-sharing, but 
may instead have a bridging tetrahedron between them. There are two forms for the 60-atom ring of 
six face-sharing icosahedra, which are analogous to the boat and chair forms 
of cyclohexane. The boat form has a slightly lower energy (by 0.391) because
a nearest-neighbour contact can be formed between the two icosahedra above the ring.

At larger sizes three-dimensional network structures begin to form. 
For example, the 67-atom global minimum involves two perpendicular interlocking planar units, 
and the 86-atom structure is a three-dimensional network of eight, mainly face-sharing, 
icosahedra, with a complex pore structure 
(the six holes on the surface of the cluster connect to a single central pore).
The formation of this latter three-dimensional structure leads to a substantial 
increase in stability (Figure \ref{fig:EvN}).
Above $N$=86 similar structures dominate the lowest-energy minima that we have found.

From these results we conclude that Dzugutov clusters tend to form non-compact polytetrahedral structures,
based on packings of 13-atom icosahedra.
Exceptions to this trend occur at $N$=57 and 76, 
where the global minima are high-symmetry compact polytetrahedral structures
involving an ordered array of disclinations. 
As a consequence of this compactness these structures have a much higher strain energy than other global minima of similar size.
They are also particularly stable for a modified Dzugutov potential that 
has been designed to favour compact polytetrahedral clusters by having a lower value of $n_{\rm eff}$,\cite{Doye00f}
and were only found on reoptimization of the structures obtained with this new potential.
The structure of these new clusters will be described in detail elsewhere.\cite{Doye00f}

Although the present work indicates that the lowest energy structures of small clusters 
are non-compact and polytetrahedral, a thorough study of the relative stabilities of possible bulk structures 
strongly suggests that for bulk a body-centred cubic (bcc) crystal is lowest in energy.\cite{Roth00}
Therefore, at sufficiently large sizes the lowest-energy clusters must be bcc.
A simple comparison of $E_{\rm ave}$, a polynomial fit to the energy of the global
minimum, with a similar polynomial function fitted to the energies of bcc rhombic dodecahedral clusters 
suggests that bcc clusters only become lower in energy beyond $N\sim 10^5$.

\begin{table*}
\caption{\label{table:dz13}
Comparison of the Dz$_{13}$ and LJ$_{13}$ PESs.
The samples of stationary points contain $n_{\rm min}$ minima and $n_{\rm ts}$ transition states. 
For the $n_{\rm search}$ lowest-energy minima transition state searches 
were performed parallel and anti-parallel to the Hessian eigenvectors with the $n_{\rm ev}$ lowest eigenvalues.
Of the transition states found, $n_{\rm ts}^{I_h}$ were connected to the global minimum.
$\left<\nu_{\rm min}\right>$ is the average geometric mean frequency of a minimum and
$\left<\nu_{\rm ts}^{\rm im}\right>$ is the average magnitude of the imaginary frequency of a transition state.
$\Delta E$, $b_u$ and $b_d$ are the energy difference, uphill barrier and downhill barrier, respectively,
between two minima connected by a single transition state.
$S$ is the integrated length of the pathway between the two minima and $D$ is the distance between the two minima.
$\tilde{N}$, the cooperativity index, gives a measure of the number of atoms involved in the rearrangement.\cite{WalesDMMW00,StillW83a}
The average values of these quantities over all the non-degenerate rearrangement pathways are given below.
(A degenerate pathway connects different permutational isomers of the same minimum.)
$\left<\Delta E\right>=\left<b_u\right> -\left<b_d\right>$.
The units of energy and frequency are $\epsilon$ and $(\epsilon/r_{\rm eq}^2)^{1/2}$, respectively,
facilitating comparison with results for 13-atom Morse clusters.\cite{Miller99a}
}
\begin{ruledtabular}
\begin{tabular}{cccccccccccccc}
 & $n_{\rm min}$ & $n_{\rm ts}$ & $n_{\rm ev}$ & $n_{\rm search}$ & $n_{\rm ts}^{I_h}$ & 
 $\left<\nu_{\rm min}\right>$ & $\left<\nu_{\rm ts}^{\rm im}\right>$ &
 $\left<\Delta E\right>$ & $\left<b_{u}\right>$ & $\left<b_{d}\right>$ &
 $\left<S\right>$ & $\left<D\right>$ & $\left<\tilde{N}\right>$\\
\hline
 LJ$_{13}$ &  $1467$ & $12\,435$ & 15 &  $1467$ & 894 & 1.671 & 0.464 & 1.593 & 2.201 & 0.609 & 
1.960 & 1.201 & 5.79 \\
 Dz$_{13}$ & $28\,568$ & $52\,813$ &  6 & $13\,675$ &   9 & 1.842 & 0.840 & 1.570 & 2.604 & 1.034 & 
1.520 & 0.958 & 3.61 \\
\end{tabular}
\end{ruledtabular}
\end{table*}

For Dzugutov clusters, structures with disclination lines, although not dominant, can be
competitive at some sizes.
Similarly for bulk, under certain conditions (when the system is somewhat compressed)
a Frank-Kasper $\sigma$-phase is lowest in energy.\cite{Roth00}

In recent simulations of a bulk Dzugutov supercooled liquid, domains
corresponding to extended icosahedral configurations have been found to 
grow as the temperature is decreased and lower-energy minima are sampled.\cite{Dzugutov00}
Much lower diffusion rates are associated with atoms in these domains compared with atoms in the rest of the liquid. 
The configurations of these icosahedral domains look very similar to the types of non-compact
polytetrahedral clusters that we observe for clusters. 
Both the structures of the clusters and the domains have a common origin in the form of the potential. 
In particular, the potential maximum causes polytetrahedral structures to be favoured,
and of the possible polytetrahedral structures the width of the potential favours 
non-compact forms based on icosahedra.

Our results also indicate that high-density, compact polytetrahedral structures are disfavoured
because they are associated with excessive values of $E_{\rm strain}$ and $E_{\rm nnn}$. 
This result has implications for the form of the phase diagram, and in particular the stability of a dense liquid.
Most simulations of Dzugutov liquids have been performed at constant volume. However, if a Dzugutov
crystal is heated at a low constant pressure then there is a transition temperature at which the system 
jumps straight to a low-density phase that resembles a foam.\cite{Roth00b} 
Further simulations suggested that there is no region of the phase diagram where a liquid is stable.\cite{Roth00b} 

Much work has been done on the effect of the range of a potential on the phase diagram. 
This research has shown that when the range is sufficiently short there 
is no stable liquid phase.\cite{Hagen93,Hagen94} The physical basis for this effect is that
as the range of the potential decreases, the potential well becomes narrower and so there is 
an increasing energetic penalty (i.e.\ $E_{\rm strain}$) for the dispersion of 
nearest-neighbour distances associated with the liquid. 
This destabilization of the liquid eventually leads to its disappearance from the 
phase diagram.\cite{Doye96a,Doye96b}

Roth attributed the absence of a liquid phase in the Dzugutov phase diagram to the range of the potential.\cite{Roth00b} 
However, the Morse potential with a similar width (i.e.\ $\rho=\rho_{\rm eff}^{\rm Dz}$) has a stable liquid phase.
Only at much larger values of $\rho$ (probably 13-14) does the liquid phase for the Morse potential become unstable.
Instead, the present study clearly indicates that the absence of a dense liquid phase for the 
Dzugutov potential stems from the maximum neart $\sqrt 2\, r_{\rm eq}$, because there are insufficient 
dense, non-crystalline structures with low values of both $E_{\rm nnn}$ and $E_{\rm strain}$.

Recently, Sear and Gelbart showed that adding a long-range repulsion of sufficient magnitude also leads to the
disappearance of the liquid from the phase diagram.\cite{Sear99a} 
As with the much shorter-ranged Dzugutov potential the positive value of $E_{\rm nnn}$ destabilizes the liquid.
Furthermore, as envisaged by Lebowitz and Penrose,\cite{Lebowitz66} 
Sear and Gilbert found that microphase separation instead occurred.\cite{Sear99a} 
Similar behaviour is seen for the Dzugutov potential:
low-density non-compact polytetrahedral structures are favourable for clusters, 
and Roth's heating simulations led to a low-density foam rather than a simple vapour.\cite{Roth00b} 
Given this possibility of low-density disordered network phases,
calculation of the true phase diagram for the Dzugutov potential is of obvious interest. 

\subsection{Energy Landscapes}

For simple pair potentials, such as the Morse and LJ forms, 
the topography of the PES for 13-atom clusters has been examined in detail,\cite{Doye00c,Miller99a}
using near complete sets of minima and transition states.
Here we examine some of the consequences of the Dzugutov potential for the topography
of the energy landscape by comparing the Dz$_{13}$ and LJ$_{13}$ PESs. 

The properties of sets of stationary points for the two clusters are 
compared in Table \ref{table:dz13}. One of the main differences between the two systems
is the much greater number of minima for Dz$_{13}$. 
We obtained roughly twenty times more minima for Dz$_{13}$, yet the sample is  
far from complete. Transition state searches have been performed from less 
than 50\% of the minima and far fewer searches were performed from each minimum than for LJ$_{13}$.

This greater number of minima is a consequence of the local maximum in the potential,
which introduces barriers between configurations that would have been in 
the same basin of attraction for LJ$_{13}$. In particular, there is now a barrier to the 
formation of nearest-neighbour contacts and so many Dz$_{13}$ minima involve atoms with low coordination numbers. 
For the LJ potential these atoms would be attracted by the long-range tail of the potential and
so would increase their coordination.
For example, when the Dz$_{13}$ minima are reoptimized for the LJ potential 502 of them
converge to the icosahedral global minimum. These differences are particularly evident in the 
connectivity of the low-energy regions of the PES. 
For LJ$_{13}$ there are over 90 times more transition states connected to the global minimum.

There are less dramatic differences between the other properties of the two PESs.
The vibrational frequencies for the Dzugutov cluster are somewhat higher because of 
its slightly larger $n_{\rm eff}$ value.\cite{Miller99a} 
For larger clusters this effect would be reinforced by the constraints on the 
second nearest-neighbour distances stemming from the maximum in the potential.
However, the latter effect is small for the 13-atom cluster because there are relatively
few second nearest-neighbours.

Barriers for the Dzugutov cluster are generally higher because the barrier for breaking
a nearest-neighbour contact involves both loss of the contact energy and overcoming the potential maximum.
In general, the barriers are also narrower and connect minima that are closer 
in configuration space, as expected for a PES with more minima.  
The rearrangements are also slightly less cooperative.
In summary, the Dz$_{13}$ PES is much rougher than the LJ$_{13}$ PES.

For larger clusters, there are generally a number of low-energy configurations
associated with different arrangements of the icosahedral units that make up most low-energy clusters. 
Here we examine the effects of such competing structures on the PES topography, 
by calculating the disconnectivity graph for a 37-atom Dzugutov cluster. 
For this cluster the global minimum, a linear quintuple icosahedron, is slightly lower in 
energy than a minimum based on the disk-like Dz$_{38}$ global minimum with an atom removed.
Next in energy above these two minima are a series of bent arrangements of five interpenetrating 
icosahedra.

Disconnectivity graphs provide a representation of the barriers between minima 
on a PES.\cite{Becker97,WalesMW98,WalesDMMW00}
In a disconnectivity graph, each line ends at the energy of a minimum.
At a series of equally-spaced energy levels we determine which (sets of) minima are connected
by paths that never exceed that energy.
Lines in the disconnectivity graph join at the energy level where
the corresponding (sets of) minima first become connected.
In a disconnectivity graph an ideal single-funnel\cite{Leopold,Bryngelson95} PES is represented by a
single dominant stem associated with the global minimum to which the other minima
join directly. For a multiple-funnel PES there are a number of major stems which only join
at high energy.

\begin{figure}
\includegraphics[width=8.2cm]{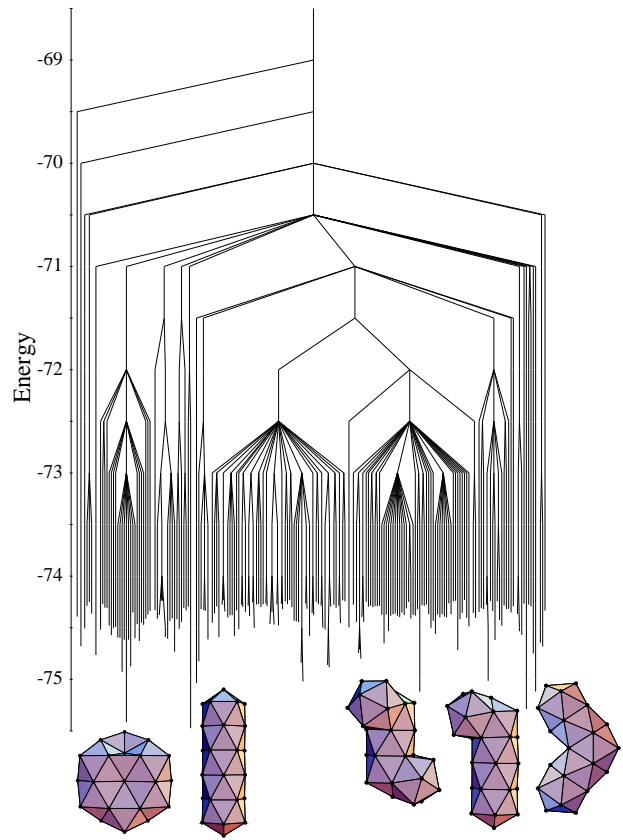}
\caption{\label{fig:tree} 
Disconnectivity graph for Dz$_{37}$. 
Only branches leading to the 200 lowest-energy connected minima are included.
The graph was calculated from a set of $30\,000$ minima and $33\,520$ transition states.
The five lowest-energy minima are depicted alongside the appropriate branches of the graph.
}
\end{figure}

The Dz$_{37}$ disconnectivity graph clearly has a multiple-funnel character. 
Each of the five lowest-energy structures are separated by large barriers, and
the corresponding branches all join up at the node at $E$=$-70.5$.
For example, the barriers for the lowest-energy pathway between the funnels of the two 
lowest-energy minima are $8.43\epsilon$ and $8.33\epsilon$. 
These large barriers will strongly influence the dynamics, making
the rate of interfunnel passage extremely slow. 
Hence, the low-energy minima act as efficient kinetic traps.

The results for the two clusters in this section also show why the global minima of 
Dzugutov clusters are particularly difficult to locate. 
Both PES roughness\cite{Miller99b} and multiple-funnel PES topography\cite{Doye98a,Doye98e,Doye99f,Doye00e} 
have been shown to hinder global optimization. The latter is particularly troublesome, when, as with Dz$_{37}$,
the global minimum is at the bottom of a narrow funnel and is thermodynamically most stable 
only at low temperatures.

These results also have implications for the dynamics of the supercooled Dzugutov liquid.
Low-energy liquid configurations with different extended icosahedral domains\cite{Dzugutov00} will be separated by
large energy barriers in a manner similar to Dz$_{37}$. Therefore, as the average energy of the liquid
minima decreases with temperature and the icosahedral domains grow, we expect
the effective activation energy associated with diffusion or viscosity to increase, 
thus leading to fragile (non-Arrhenius) dynamics.\cite{Dzugutov91,Dzugutov00} 

\subsection{Conclusions}

The global minima of Dzugutov clusters have non-compact polytetrahedral structures.
The interior atoms in these clusters almost always have an icosahedral coordination shell, 
rather than other polytetrahedral coordination shells that involve disclination lines,
because the latter involve too large a strain energy. 
The 13-atom icosahedral units that make up the global minima are usually interpenetrating or face-sharing,
and both these ways of linking icosahedra are commonly observed in metallic crystals that have icosahedral 
coordination.\cite{Shoemaker}
As the size of the clusters increases the arrangements of the icosahedral units 
change from one-dimensional chains to two-dimensional rings to three-dimensional porous networks.
These results are consistent with the previous observation that the bulk liquid can sustain
a distribution of free volume in the form of atomic size vacancies.\cite{Sadigh99}

Unfortunately, Dzugutov clusters, because of their non-compact character, do not provide
a realistic model for polytetrahedral metallic clusters, such as those recently 
observed for cobalt.\cite{Dassenoy00} However, promising results have been obtained from a modified version of 
the Dzugutov potential that is better able to accommodate the strain associated with compact
polytetrahedral structures. Results for this system will be reported elsewhere.\cite{Doye00f}

Dzugutov clusters typically have rough multiple-funnel energy landscapes. These features makes global optimization
difficult and have implications for the dynamics of supercooled Dzugutov liquids, where extended 
icosahedral domains,\cite{Dzugutov00} which are similar to the structures we find for the clusters, 
grow as the temperature is decreased. 
Furthermore, our results help to explain why the bulk phase diagram appears not to include a stable liquid,
and suggest the possibility of phases with complex low-density network structures.

\begin{acknowledgments}
J.P.K.D. is the Sir Alan Wilson Research Fellow at Emmanuel College, Cambridge.
We gratefully acknowledge helpful discussions with Mikhail Dzugutov and Richard Sear. 
\end{acknowledgments}


\begin{thebibliography}{10}
\bibitem{Frankbug}
\bibinfo{note}{There is actually an error in Frank's argument. For clusters,
  the icosahedron is energetically favoured over a close-packed configuration
  because it has a lower surface energy. Such a surface effect cannot be
  relevant to bulk liquids, and so the origin of local icosahedral order in
  liquids is more complex.}
\bibitem{ptet}
\bibinfo{note}{For a polytetrahedral structure all of space can naturally be
  divided up into tetrahedra.}
\end{thebibliography}
\end{document}